\documentclass[11pt]{article}
\usepackage[a4paper, margin=2cm]{geometry}
\usepackage{graphicx} 
\usepackage{url}
\usepackage[x11names]{xcolor}
\usepackage{tabularx}
\usepackage{hyperref}
\usepackage[super, numbers, sort]{natbib}
\usepackage{caption}
\usepackage{enumitem}
\colorlet{shadecolor}{LavenderBlush3}
\usepackage{framed}
\usepackage{comment}
\usepackage{wrapfig}
\usepackage{multicol}

\setlist[itemize,1]{leftmargin=\dimexpr 15pt}

\usepackage{etoolbox}
\patchcmd{\thebibliography}{\clubpenalty4000}{\clubpenalty4000\setlength{\itemsep}{0pt plus 1pt minus 1pt}}{}{}
\patchcmd{\thebibliography}{\widowpenalty4000}{\widowpenalty4000\setlength{\itemsep}{0pt plus 1pt minus 1pt}}{}{}

\title{{\textbf{Magnetohydrodynamic seismology of \\ solar and heliospheric plasmas}}\\
{\textit{Thematic Area: the Sun and Heliosphere (Helio)}}}

\vspace{-5mm}
\author{{\textit{Lead Author:}} V.M. Nakariakov (\href{mailto:V.Nakariakov@warwick.ac.uk}{V.Nakariakov@warwick.ac.uk}) \\ D.B. Jess, A.N. Wright, T.K. Yeoman, T. Elsden,  \\ J.A. McLaughlin, D.Y. Kolotkov, V. Fedun \& R. Erd\'elyi}

\date{November 2025}

\begin{document}

\maketitle
\setcounter{page}{1}
\thispagestyle{empty}

\vspace{-10mm}
\begin{center}
\includegraphics[width=0.99\textwidth]{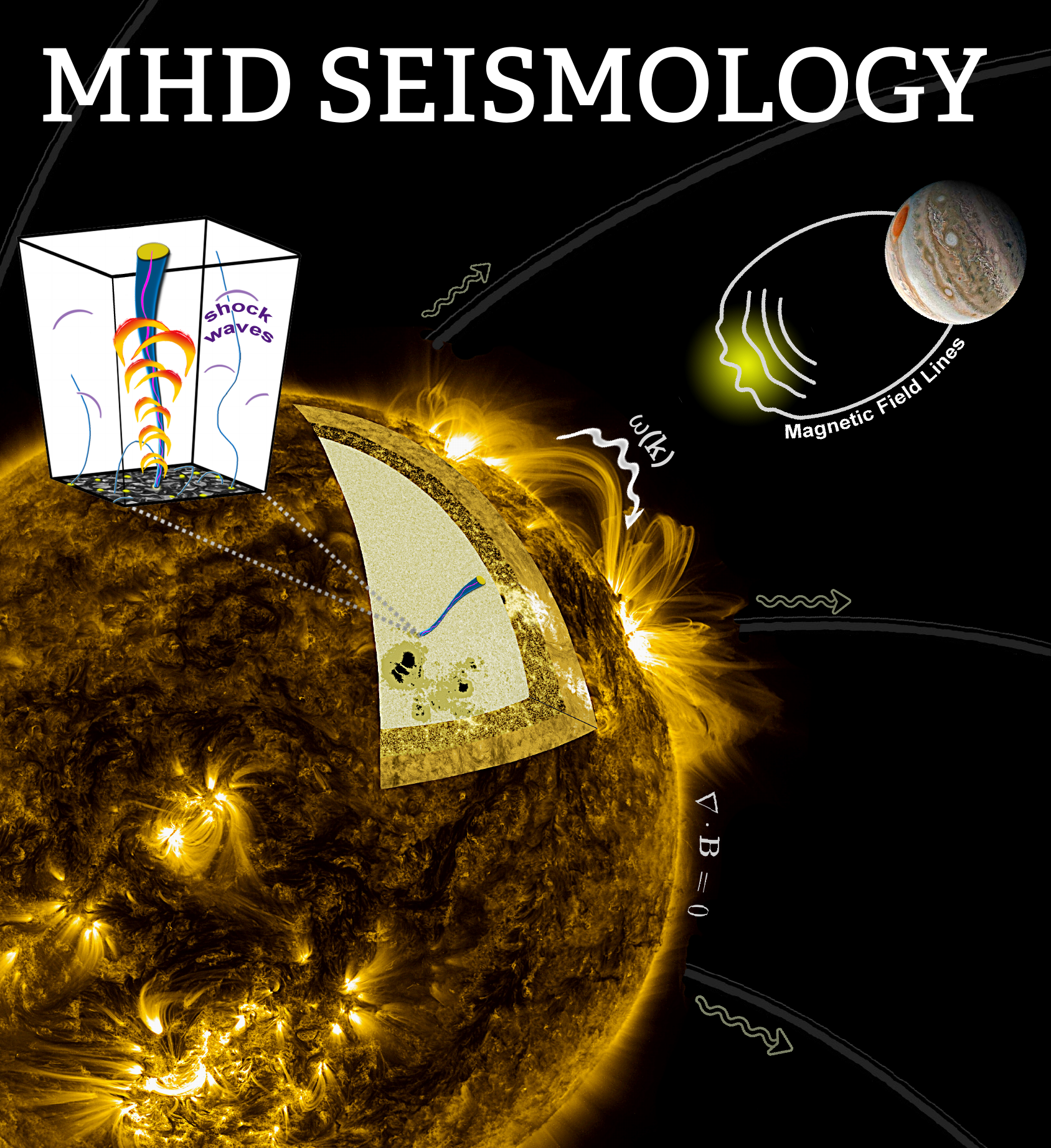}
\end{center}

\vspace{-6.0mm}
\begin{flushright}
{\textit{Image credit: Dr Shahin Jafarzadeh (QUB) and the NASA instrument teams}}
\end{flushright}

\newpage
\setcounter{page}{1}
\begin{center}
    {\fontsize{12pt}{16pt} \selectfont \textbf{Magnetohydrodynamic seismology of solar and heliospheric plasmas}} \\
V.M. Nakariakov, D.B. Jess, A.N. Wright, T.K. Yeoman, \\ T. Elsden, J.A. McLaughlin, D.Y. Kolotkov, V. Fedun,  \& R. Erd{\'{e}}lyi
\end{center} 

~ \\

\vspace{-10mm}
\noindent\rule[7pt]{\linewidth}{0.4pt}

\vspace{-8.3mm}
\begin{snugshade*}
\begin{center}
{\bfseries Scientific motivation and objectives} 
\end{center}
\end{snugshade*}

\vspace{-5.3mm}
\noindent\rule[7pt]{\linewidth}{0.4pt}

\vspace{-3mm}
The solution to major open questions in solar and heliospheric physics, e.g., the mechanisms behind chromospheric and coronal heating, flares and eruptions, solar wind acceleration, space-weather effects, and planetary magnetospheric dynamics, requires reliable estimates of plasma conditions and compositions. However, key variables such as magnetic fields, currents, fine transverse structuring, transport coefficients, energetics, and connectivity are difficult or impossible to measure directly. Thankfully, a powerful indirect diagnostic technique relies on the study of magnetohydrodynamic (MHD) waves supported by these plasma structures. ``MHD'' reflects the long wavelengths and periods involved, from tens to hundreds of thousands of km and from seconds to hours, far exceeding microscopic plasma scales \cite{2022STP.....8d...3C}.

\vspace{1.5mm}
\hspace{5mm} Throughout the outer solar atmosphere MHD waves are detected as propagating or standing periodic disturbances in UV, EUV, soft X-ray, visible radiation, and spectroscopic line variations\cite{2020ARA&A..58..441N, 2023LRSP...20....1J} (see Figure~{\ref{fig:waves_stack}}). They modulate radio and microwave bursts and are measured {\textit{in-situ}} at all magnetised planets. Theory links observable properties to plasma structures, enabling MHD seismology\cite{2024RvMPP...8...19N}. Dynamic atmospheres are present in other cool, Sun-like stars, whose dynamic processes may affect habitability or even support prebiotic conditions\cite{2018SciA....4.3302R}, motivating robust diagnostic techniques based on the solar-stellar analogy.

\vspace{1.5mm}
\hspace{5mm} From the photosphere to the corona, MHD waves propagate within diverse waveguides (including sunspots\cite{2017ApJ...842...59J, 2018NatPh..14..480G, 2021ApJ...914L..16C, 2023ApJ...954...30A}, pores\cite{2011ApJ...729L..18M, 2015ApJ...806..132G, 2018ApJ...857...28K, 2021RSPTA.37900172G, 2021NatAs...5..691S}, loops\cite{1999Sci...285..862N, 2002SoPh..209...61D,  2015A&A...583A.136A, 2025ApJ...982..202D}, jets\cite{2021ApJ...913...19M, 2022NatPh..18..595D}, spicules\cite{2009Sci...323.1582J, 2009SSRv..149..355Z, 2012ApJ...744L...5J, 2018ApJ...853...61S, 2019Sci...366..890S, 2022ApJ...930..129B}, fibrils\cite{2012NatCo...3.1315M, 2017A&A...607A..46M, 2024ApJ...970...66B}, and plumes\cite{1999ApJ...514..441O, 2025ApJ...991L..45B, 2025ApJ...992...33C}) and include fast and slow magnetoacoustic, Alfv{\'{e}}n, and entropy modes \cite{2013SSRv..175....1M, 2015SSRv..190..103J}. In addition, chromospheric waves are able to trace transitions between gas- and magnetic-pressure-dominated regimes. New and upcoming facilities, such as DKIST, SKA, Solar-C, and EST, will deliver unprecedented spatial, spectral, and temporal resolutions, demanding a deeper understanding of the Sun's underlying plasma physics.

\vspace{1.5mm}
\hspace{5mm} Kink oscillations of magnetic structures appear as transverse displacements or Doppler-shift oscillations\cite{2021SSRv..217...73N, 2023A&A...678A.205C}. In the corona, they occur in two regimes: large-amplitude, rapidly decaying oscillations and low-amplitude decayless motions, both with periods of $\sim$1--30~min. Their phase structure and period-length scales confirm their standing-wave nature. Decaying and decayless oscillations allow us to probe magnetic fields, electric currents, transverse structuring, and the coronal energy supply. Further down in the solar chromosphere, narrow magnetic structures such as spicules, fibrils, and mottles, increasingly show high-frequency, high-amplitude transverse wave activity \cite{2022ApJ...930..129B}. These rapidly evolving features, visible only in the highest-resolution datasets, may illuminate mass and energy transport throughout the outer solar atmosphere. Continued development of advanced instrumentation sensitive to fast dynamics and weak spectropolarimetric signals is therefore essential for improving our understanding of wave energetics and energy coupling within the lower solar atmosphere.

\vspace{1.5mm}
\hspace{5mm} Propagating slow waves manifest as nearly monochromatic intensity disturbances guided upward from the lower solar atmosphere\cite{2021SSRv..217...76B}, typically with $\sim$3-min periods and several hundred km/s speeds in the corona. They damp strongly with height and diagnose magnetic-field orientations and thermal structuring\cite{2014A&A...561A..19Y, 2016NatPh..12..179J}. Chromospheric signatures are often visualised through phenomena including running penumbral waves\cite{1972ApJ...178L..85Z, 2007ApJ...671.1005B, 2013ApJ...779..168J, 2014ApJ...791...61F} and umbral flashes \cite{1969SoPh....7..351B, 2013A&A...556A.115D, 2020A&A...642A.215H}, providing insight into non-linear wave dynamics \cite{2020ApJ...892...49H, 2021A&A...645A..81S}. In the corona, slow-wave ``sloshing'' in hot ($>$6~MK) loops\cite{2007ApJ...656..598W, 2021SSRv..217...34W} shows plasma bouncing between footpoints with periods $>$10~min and rapid damping, informing thermal conduction and coronal heating. Quasi-periodic fast propagating (QFP) waves appear as EUV ripples with speeds of several hundred km/s and $\sim$1-min periods\cite{2022SoPh..297...20S}. Their magnetoacoustic nature makes them sensitive to magnetic geometry and plasma non-uniformity, although their seismological use is still emerging.

\vspace{1.5mm}
\hspace{5mm} Quasiperiodic pulsations (QPPs) during magnetic reconnection show repetitive modulations of emission\cite{2018SSRv..214...45M}, with periods from fractions of a second to tens of minutes. Understanding QPP mechanisms provides insight into flare energy release and supports the solar–stellar analogy.

\vspace{-3mm}
\begin{minipage}[!h]{0.39\textwidth}
\vspace{0mm}
\hspace{5mm} Strong stratification in the lower solar atmosphere highlights several key MHD processes. Non-linear shock formation is a potential heating mechanism\cite{2018ApJ...860...28H, 2022NatPh..18..595D}, while modern theory\cite{2023MNRAS.525.4717S} clarifies ion-neutral effects. Large-scale wave eigenmode reflections, especially in sunspot umbrae, reveal possible MHD resonance cavities\cite{2011ApJ...728...84B, 2020NatAs...4..220J}. Their interaction with non-linear shocks adds complexity that requires instruments capable of detecting small-scale magnetic fluctuations, which is an area with strong UK heritage (e.g., Hinode/EIS, DST/ROSA, Solar Orbiter/SPICE).

\vspace{2mm}
\hspace{5mm} Ultra-Low Frequency (ULF) waves are the lowest frequency wave modes of the Earth's magnetosphere and are commonly studied using MHD \cite{2024GMS...285..215W}. This setting permits {\textit{in-situ}} observation of the waves, so is complementary to the solar environment that enables remote observations of the global system. ULF waves play a key role in transporting energy and momentum, which originate in the solar wind, through to the magnetosphere and ionosphere. Magnetoseismology is best developed in the magnetosphere of the Earth, where, for example, it has been possible to use magnetic 
\end{minipage} \hfill
\begin{minipage}[!h]{0.59\textwidth}
\vspace{0mm}
\centering
\includegraphics[width=\textwidth]{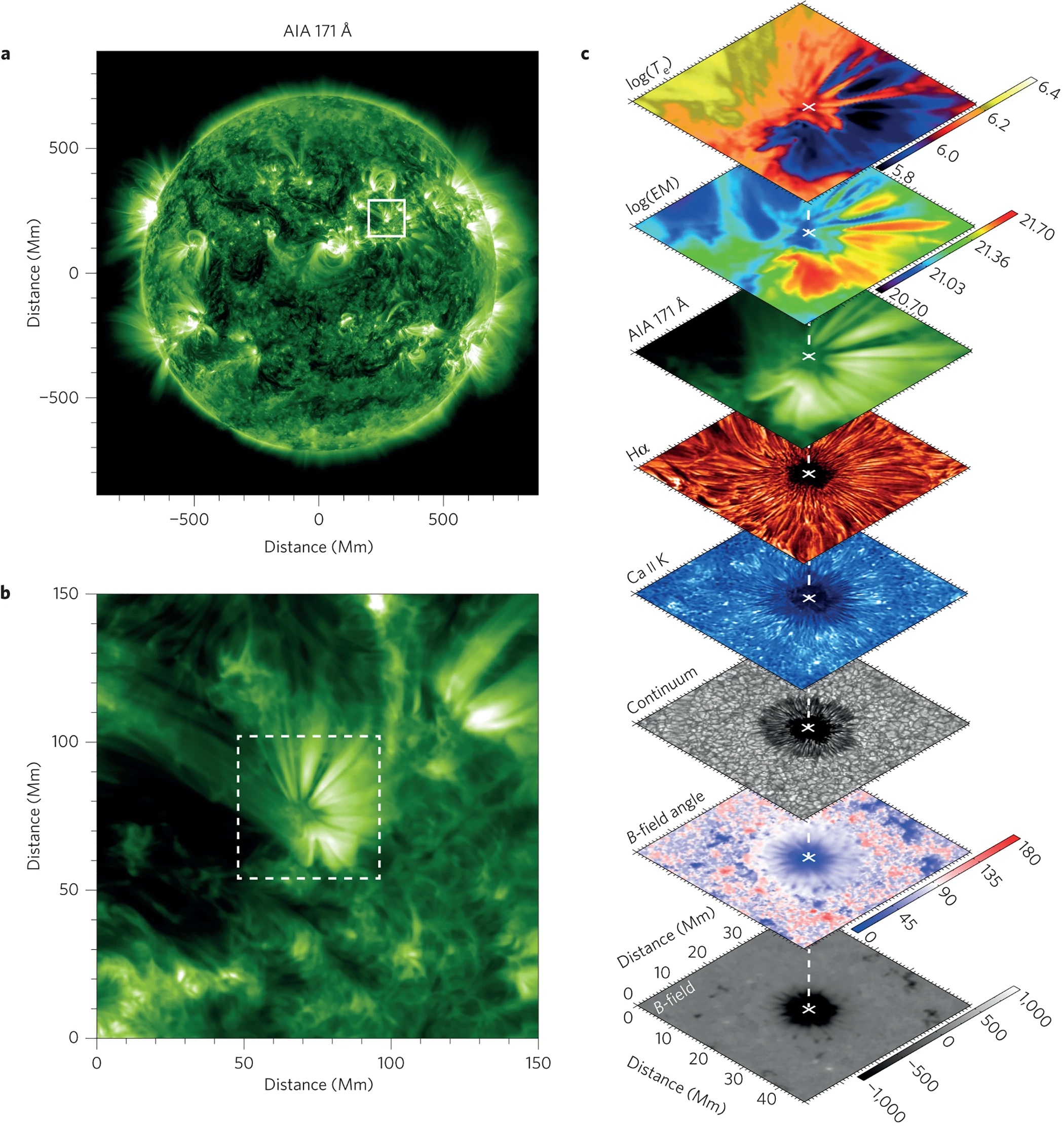}

\vspace{-1mm}
\captionof{figure}{{\color{gray}{Multiwavelength demonstration of the magnetic coupling between various layers of the solar atmosphere\cite{2016NatPh..12..179J}. Panels {\textbf{a}} \& {\textbf{b}} showcase the Sun's corona, where magnetic loops and fans are seen to oscillate with a variety of MHD wave modes. Panel~{\textbf{c}} reveals the atmospheric coupling, from the photosphere to the corona, that supports and guides wave energy into the heliosphere and beyond. }}}
\label{fig:waves_stack}
\end{minipage}

\vspace{1mm}
field measurements during magnetic storm periods to demonstrate up to 50\% changes in field line eigenfrequencies\cite{2020JGRA..12527648W}, hence diagnosing changing magnetic fields and plasma mass densities.

\vspace{2mm}
\hspace{5mm} At other planets, where typically data are only available from single spacecraft and complementary ground-based assets are unavailable, magnetoseismology can still offer profound insights into magnetospheric dynamics, e.g., from investigations of periodic features in the planetary auroras \cite{2017GeoRL..44.9192N}. Such investigations of the auroras of Jupiter and Saturn have revealed new insights into the magnetospheric morphology and the dynamics of hot plasma populations.  At Mercury, statistical studies of ULF waves in the in-situ magnetic field have provided detailed models of the density and composition of the magnetospheric plasma and the morphology of the highly dynamic planetary magnetic field \cite{2019JGRA..124..211J}.

\vspace{2mm}
\hspace{5mm} The full-scale application of MHD seismology to the outer solar atmosphere and magnetospheres is made challenging by several factors. In the solar atmosphere, one of the key challenges is the lack of reliable diagnostics of plasma density and temperature, which are typically inferred from either emission produced by minor species or the outputs from spectropolarimetric inversions. This issue could be mitigated through radio and microwave observations; however, the relatively low spatial resolution of such data may prevent the emission from being associated with specific plasma structures. The identification of particular MHD modes, or combinations of coupled modes, would be greatly improved if high-resolution imaging and spectral information were available simultaneously along the same plasma structure. Several seismological techniques require stereoscopic, or at least quasi-stereoscopic, observations from different (and ideally non-coplanar) lines of sight. The detection of higher harmonics, including signatures of non-linear cascades and the seismological exploitation of their properties, is limited by the insufficient resolution of current instruments. 

\newpage
\hspace{5mm} Theoretically, there is a need to understand and minimise numerical artefacts, as well as to achieve a more effective integration of HPC and analytical modelling. Furthermore, in chromospheric wave modelling, there is a clear need to account for the effects of partial ionisation and non-local thermodynamic equilibrium, while in the corona, collisionless and non-adiabatic effects on MHD waves must be better understood and incorporated into modelling efforts. Within magnetospheric contexts, maintaining dense arrays of ground magnetic observatories on Earth remains a challenge. The scientific exploitation of new, fully-digital radar techniques is in its very early stages, but offers immense promise for the improvement of temporal and spatial resolutions in data collection. Instrumentation is non-viable at the surface of other planets. However, a multispacecraft era beckons and continuous improvements in remote sensing techniques, e.g., in the IR, UV and soft X-ray wavelengths, offer future opportunities to image boundaries and aurora from both orbiting planetary  spacecraft and vantage points such as at JWST. 

~ \\

\vspace{-4.2mm}
\noindent\rule[7pt]{\linewidth}{0.4pt}

\vspace{-8.3mm}
\begin{snugshade*}
\begin{center}
{\bfseries Strategic context} 
\end{center}
\end{snugshade*}

\vspace{-5.3mm}
\noindent\rule[7pt]{\linewidth}{0.4pt}

\vspace{-2mm}
Reliable diagnostics of solar plasma environments using MHD seismology would, in general, enhance our understanding of the fundamental physical processes operating throughout the outer solar atmosphere (photosphere to corona), the Earth's and planetary magnetospheres, and in the solar wind, which can then be incorporated into space weather forecasting models. In addition, advanced seismology will supply these models with invaluable input parameters. Variations in oscillation parameters may also indicate sub-critical states of active regions, and thus act as precursors to potentially geoeffective flares and eruptions \cite{2022ApJ...933...66K}. These advances are expected to significantly improve space weather forecasts, providing substantial benefits for UK industries, including telecommunications, aviation, power supply, and other stakeholders.

\vspace{1mm}
\hspace{5mm} Novel plasma diagnostic techniques may also be transferred to laboratory plasma research, including controlled fusion experiments, which are of paramount importance to achieve future fossil-fuel-free energy production. Furthermore, emerging data analysis tools, such as methods for analysing non-stationary and non-linear oscillatory processes\cite{2022SSRv..218....9A, 2025NRvMP...5...21J}, and machine learning and AI solutions required by MHD seismology, are of growing interest beyond solar physics, with potential applications spanning engineering, medicine, security, climate research, geophysics, and econometrics.

~ \\

\vspace{-4.8mm}
\noindent\rule[7pt]{\linewidth}{0.4pt}

\vspace{-8.3mm}
\begin{snugshade*}
\begin{center}
{\bfseries Proposed approach} 
\end{center}
\end{snugshade*}

\vspace{-5.3mm}
\noindent\rule[7pt]{\linewidth}{0.4pt}

\vspace{-3mm}
The very nature of MHD seismology is the synthesis of high-precision observations, advanced and fine-tuned data analysis techniques, with state-of-the-art theoretical modelling.

\vspace{-2mm}
\subsubsection*{Observational challenges}
\vspace{-2mm} Accurate characterisation of MHD wave activity requires observations to be obtained that encapsulate the simultaneous trifecta of high spatial, spectral, and temporal resolutions. This is due to oscillatory behaviours demonstrating displacements of a few tens of km \cite{2023A&A...671A..69G}, Doppler velocities of a few tens of m/s \cite{2018A&A...617A..19L}, and periods on the order of a few seconds \cite{2022ApJ...930..129B}. This is a very challenging endeavour, since while the Sun is an inherently bright source, observations performed at the diffraction limit often suffer from photon starvation, particularly when narrowband spectral images are acquired and the incoming light is further decomposed into its constituent Stokes vectors representative of the incident polarisation states. Therefore, pushing the boundaries of MHD wave detection, extraction, and quantification requires the community to continue its development of next generation instrumentation that can provide scientists with multi-wavelength measurements of plane-of-sky displacements, Doppler motions, and spectropolarimetric signatures that are suitable for use with modern inversion routines \cite{2022A&A...660A..37R}, all while obtaining these data at the highest temporal cadences possible. Thankfully, the UK has recently demonstrated its capability to do this with the commissioning of the world's first solar integral field unit \cite{2023SoPh..298..146J} (IFU) designed for the blue portion of the optical spectrum, where many photospheric and chromospheric lines reside. Hence, there is an existing skillset to build upon within the UK to develop the next generation of instruments capable of contributing to step-change advancements in MHD seismology of the solar atmosphere.

\begin{minipage}[!h]{0.62\textwidth}
\vspace{0mm}
\hspace{5mm} Future exploitation of magneto(spheric)seismology within planetary magnetospheres will require continued multi-instrument, in-situ spacecraft observations from both terrestrial and planetary missions.  Of particular importance is the multipoint opportunities offered by spacecraft constellations and microsatellites. Novel imaging techniques, such as energetic neutral atom imaging\cite{2007RvGeo..45.4003F} and multiwavelength auroral imagery, will provide complementary datasets. At Earth, maintaining a close coordination between in-situ observations of the magnetospheric plasma and extensive, global-scale networks of ground instrumentation (e.g., SuperMAG and SuperDARN) have proven to be effective approaches, and achieving continuity of operations of these networks is essential.

\vspace{1mm}
\hspace{5mm} Promising opportunities for MHD seismology are offered by the UK involvement in the Square Kilometre Array (SKA) project. The high spatial resolution of SKA-Mid, down to a fraction of an arcsecond in optimal regimes, will enable pinpointing the fine structures of QPPs in flaring regions. Moreover, its fast imaging mode, when used in a custom observing strategy, could achieve cadences on the order of $\sim$1~s, potentially allowing the exploitation of higher harmonics, the detection of more challenging MHD modes (e.g., sausage modes), and the study of non-linear cascade processes.

\subsubsection*{State-of-the-art theoretical modelling}
\vspace{-2mm} It is widely accepted that MHD wave theory provides an accurate description of most macroscopic wave phenomena observed in the corona and some parts of the magnetosphere, and serves as a robust first-order approximation for waves detected in the photosphere/chromosphere and for large-scale disturbances throughout the heliosphere. Central to theoretical modelling is understanding how MHD waves propagate and couple in non-uniform media. One of the key processes in this context is the resonant coupling of fast MHD waves to Alfv{\'{e}}n waves \cite{2011SSRv..158..289G}. An emerging theoretical challenge is the need to treat mechanical and thermal equilibria self-consistently in chromospheric and coronal MHD wave modelling. These plasmas, undergoing continuous heating and cooling, behave as an active medium that modifies the MHD wave dynamics. In cooler regions, the additional effects of partial ionisation must also be considered\cite{2022ApJ...938..154M}.
\end{minipage} \hfill
\begin{minipage}[!h]{0.35\textwidth}
\vspace{0mm}
\centering
\includegraphics[width=\textwidth]{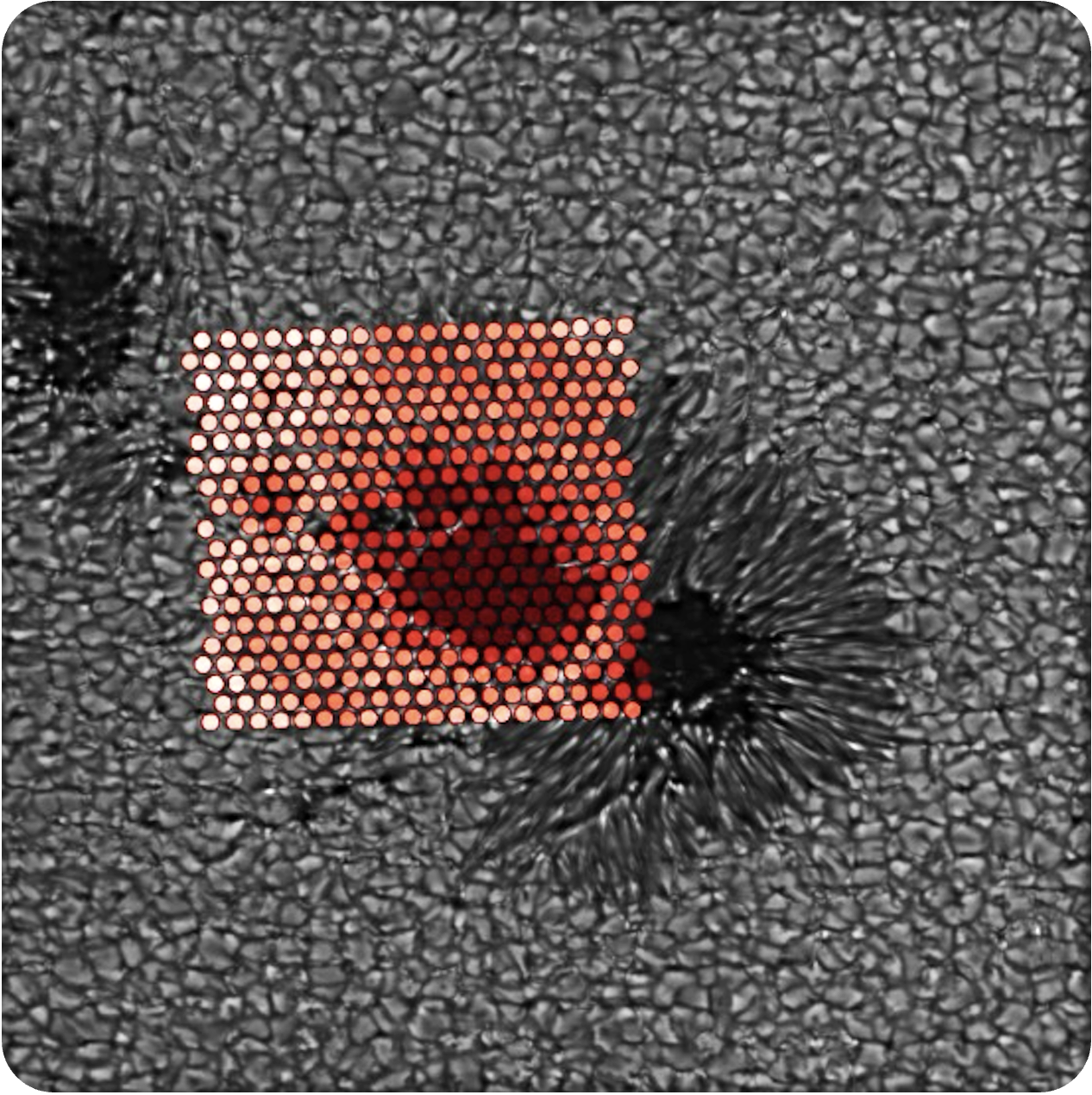}
\includegraphics[width=\textwidth]{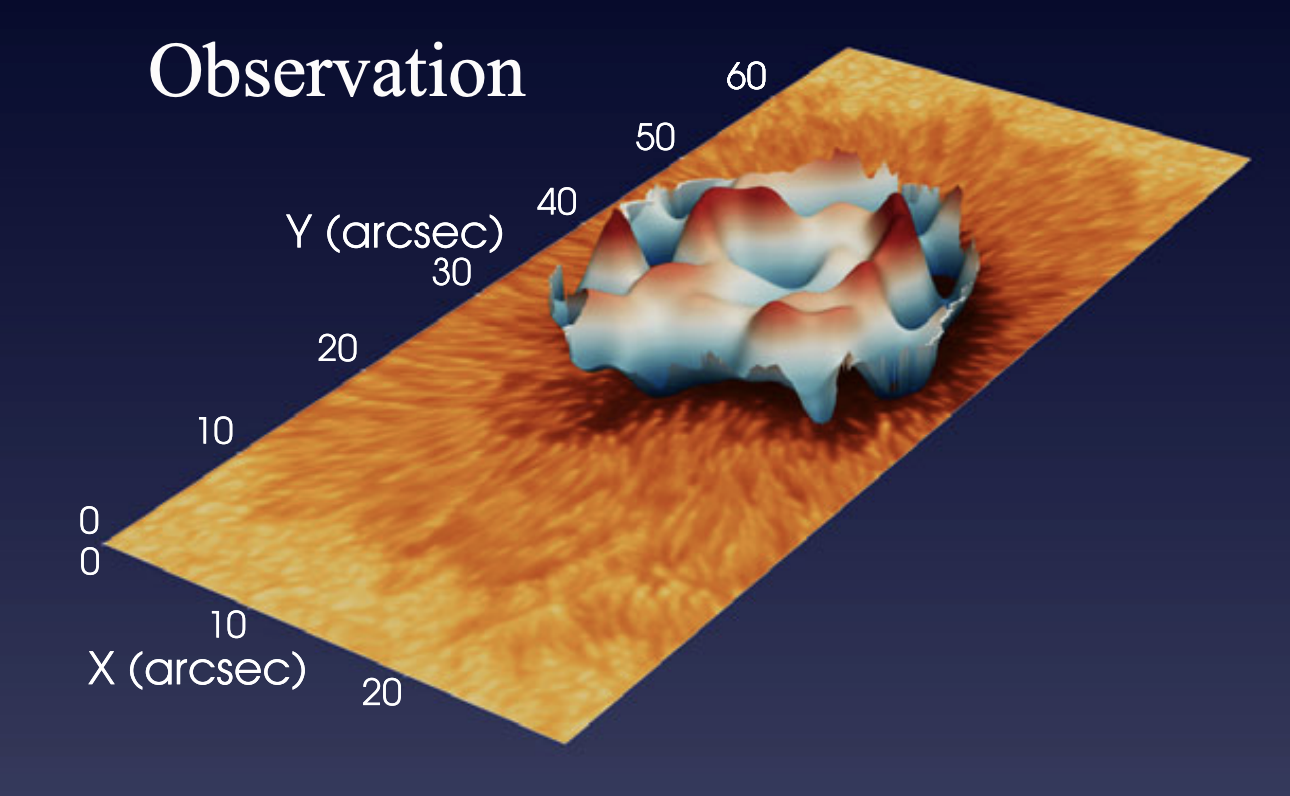}
\includegraphics[width=\textwidth]{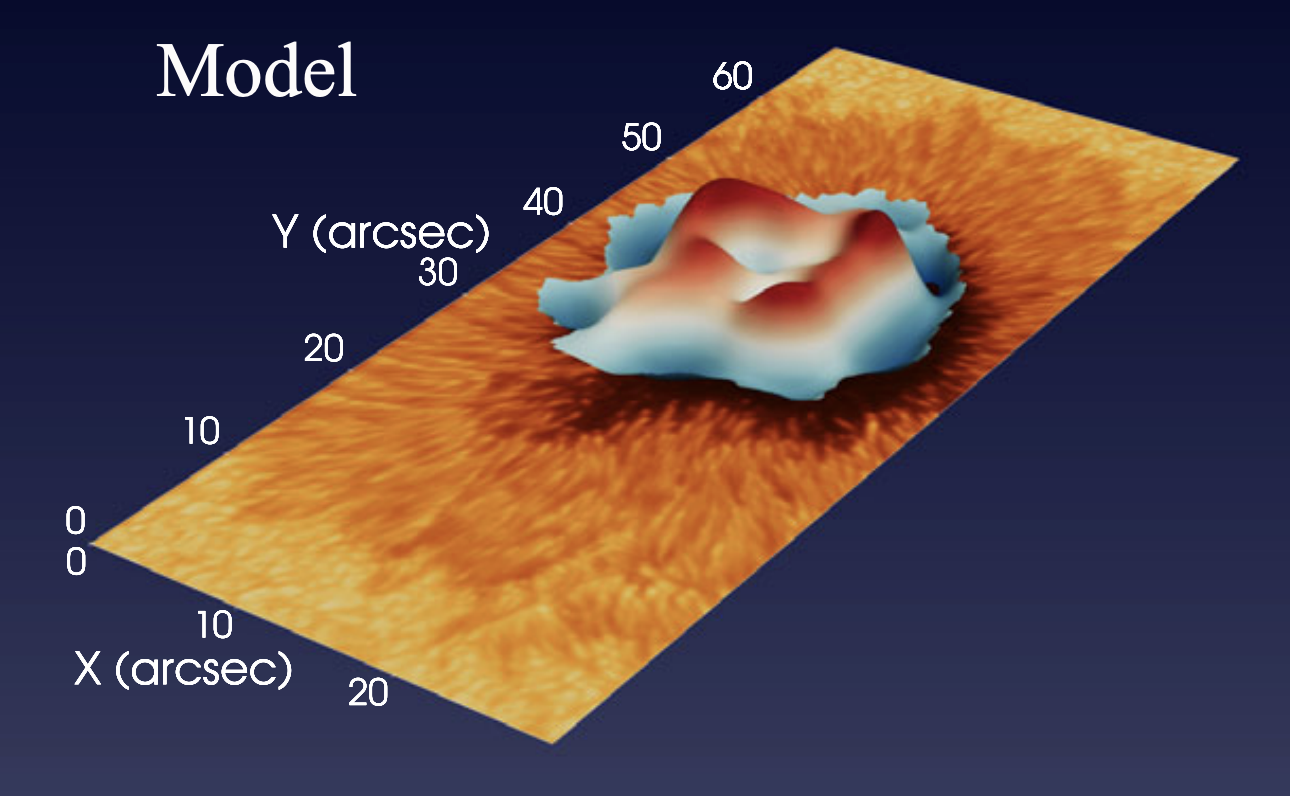}

\vspace{-1mm}
\captionof{figure}{{\color{gray}{Next-generation instrumentation\cite{2023SoPh..298..146J}, numerical modelling\cite{2023ApJ...954...30A}, and theoretical interpretation\cite{2022NatCo..13..479S} working in harmony to diagnose the modal composition of MHD waves in a sunspot umbra. Integral field units (IFUs; top panel) offer a novel solution for obtaining information across spectral and two spatial domains simultaneously to provide the high-resolution information necessary for in-depth seismological studies. }}}
\label{fig:waves_IFU}
\end{minipage}

\vspace{3.5mm}
\hspace{5mm} Other outstanding problems include modelling how MHD waves trigger magnetic reconnection, regulate energy release, and influence the acceleration and transport of non-thermal particles. Applying theoretical developments to seismological diagnostics further requires forward modelling of observational signatures as recorded by specific instruments, taking into account spatial and temporal resolution, sensitivity, response functions, and other instrumental characteristics.

\vspace{2mm}
\hspace{5mm} Progress in this field relies on the interplay between full-scale numerical simulations, which capture the complexity of real plasma systems, and advanced analytical theory, which isolates the fundamental physical mechanisms. Thankfully, with HPC systems now available within the UK for extensive computing requirements (e.g., DiRAC), it is possible for researchers to model increasingly complex MHD processes to better understand the underlying plasma physics at work in the atmospheres of the Sun and neighbouring planets. 

\subsubsection*{Advanced data analysis techniques}
\vspace{-2mm} In many cases, solar and heliospheric MHD wave processes are inherently sporadic and non-stationary, with their parameters, such as oscillation periods and amplitudes, varying on time scales comparable to the oscillation period itself. Moreover, the signals of interest are often strongly non-harmonic. As a result, traditional wave analysis techniques based on the Fourier transform are inadequate. Therefore, significant progress is expected from the use of modern data analysis methods, such as empirical and variational mode decomposition, the Hilbert transform, and their various modifications\cite{2022SSRv..218....9A, 2025NRvMP...5...21J}. Importantly, the UK has a significant track record in the development of novel wave analysis tools, many of which are available through publicly accessible code repositories to benefit the wider research community (e.g., \href{https://WaLSA.tools}{https://WaLSA.tools}). 

\subsubsection*{Machine learning and big data approaches}
\vspace{-2mm} The increasingly large volume of multi-instrument, multi-wavelength observational data motivates the use of machine learning (ML) and other big-data approaches for the automated detection of oscillatory processes, their classification, and the identification of hidden correlations and relationships \cite{2025ApJS..281...12B}. In addition, ML-based techniques can be applied to the unsupervised clustering of detected events and to the discovery of phenomena not described by existing models (i.e., anomaly detection). Moreover, combining different data sources (e.g., multi-modal ML) has the potential to substantially improve model performance and thus provide the community with fast and accurate analysis of observational data. Importantly, ML approaches have also recently been applied to high resolution spectroscopy to identify and isolate multiple blended spectral components, hence enabling higher precision characterisations (e.g., Doppler velocities, line asymmetries and widths, etc.) to be uncovered - something that will be of paramount importance for future studies involving spectropolarimetric wave signatures \cite{2021RSPTA.37900171M}.    

\hspace{5mm} Machine learning has also been harnessed in recent data reduction techniques of in-situ planetary particle data \cite{2020JGRA..12527352J} and in both the modelling of the plasma mass density of the Earth's magnetosphere and the determination of the key parameterisations required in such models \cite{2021JGRA..12629565J}.

\subsubsection*{Integration of MHD seismology with other plasma diagnostics techniques}
\vspace{-2mm} So far, seismological diagnostics of solar and heliospherical plasmas has been a stand-alone approach, and its outcomes have not been integrated with results of other diagnostic techniques. Promising further steps are the use of seismological estimates of the coronal magnetic field for constraining non-linear, force-free extrapolations of the photospheric magnetic fields and radio magnetometry. 

~ \\

\vspace{1.0mm}
\noindent\rule[7pt]{\linewidth}{0.4pt}

\vspace{-8.2mm}
\begin{snugshade*}
\begin{center}
{\bfseries Proposed technical solution and required developments} 
\end{center}
\end{snugshade*}

\vspace{-5.3mm}
\noindent\rule[7pt]{\linewidth}{0.4pt}

\vspace{-3mm}
Observing MHD wave phenomena can be undertaken using a variety of different instruments, including traditional imagers, Fabry-P{\'{e}}rot interferometers, scanning slit-based spectrographs, and radio telescopes. One drawback of these existing instrumentation suites is the fact that simultaneous measurements of spatial and spectral information are not possible, e.g., with Fabry-P{\'{e}}rot interferometers requiring to scan across a spectral line and slit-based spectrographs needing to raster to build-up two-dimensional information. This is where the novel power of integral field units (IFUs) has the potential to revolutionise the field of MHD seismology. By coupling spatial information to individual rows of spectral dispersion within a spectrograph, IFUs are able to reconstruct simultaneous three-dimensional fields of [$x, y, \lambda$], which when coupled to polarisation optics and observed at high frame rates exceeding several per second, enable revolutionary 5D cubes of [$x, y, \lambda, S, t$] to be obtained, where $S$ is the Stokes parameter of the incident polarised light. These data will deliver high-resolution maps of key neutral/ionised spectral parameters, such as Doppler shifts and non-thermal line broadening, allowing the complete evolution of the absorption/emission lines embedded across the entire solar atmosphere to be studied in unprecedented detail, particularly when combined with modern spectropolarimetric inversion algorithms \cite{2022A&A...660A..37R} to extract key plasma parameters as a function of optical depth and/or geometric height (see Figure~{\ref{fig:waves_IFU}}). 

\vspace{2mm}
\hspace{5mm} Such IFU technology has been developed in recent years within the UK \cite{2023SoPh..298..146J}, hence providing a firm foundation on which to further build and refine. The upcoming Brazilian Galileo Solar Space Telescope \cite{2024AGUFMSH11F2855V} (GSST), which has recently been promoted to the qualification portfolio of the Brazilian Space Agency, is involved in a formal agreement within the UK to develop IFU technology for space-borne satellites, hence offering global leadership potential for the UK in this timely sector. This calls for a number of challenging, yet technically achievable developments in IFU design to be driven forward by the UK community over the next decade. Notably, it will be important to design and test the new translational architecture necessary to position fibre elements within the targeted field-of-view, often with bundles of 10,000+ fibres arranged in a close hexagonal packing configuration to maximise the filling factor of the incident light, to provide the highest spatial resolutions that complement the simultaneous, high spectral resolutions. Furthermore, laboratory testing of how IFU optics degrade in challenging space-borne environments (i.e., without the protection of the Earth's atmosphere) will be an important stepping stone when finalising a space-qualified IFU. Importantly, the UK community, through UKRI-funded laboratory facilities, offers the skillsets and knowledge to combine academic expertise, a track record in instrument design, and space qualification requirements that will ensure the UK are at the forefront of large-scale space missions for decades to come. 

\hspace{5mm} Similar designs and use of space-borne instrumentation capable of delivering high-resolution maps of spectral parameters, such as Doppler shifts, non-thermal line broadening, and intensity evolution of coronal emission lines, is critical to make step-change advancements in our knowledge of MHD processes in million-degree magnetic plasmas.

~ \\

\vspace{-7mm}
\noindent\rule[7pt]{\linewidth}{0.4pt}

\vspace{-8.3mm}
\begin{snugshade*}
\begin{center}
{\bfseries UK leadership and capability} 
\end{center}
\end{snugshade*}

\vspace{-5.2mm}
\noindent\rule[7pt]{\linewidth}{0.4pt}

\vspace{-3mm}
The UK has significant expertise and world-wide leadership in magnetohydrodynamics and magnetospheric plasma seismology. Built upon the legacy of foundational work by Dungey \cite{1953MNRAS.113..180D, 1970SSRv...10..672D}, there are many institutes across the country who are actively researching MHD phenomena and developing next-generation hardware. For example, the University of Warwick led the creation of the first comprehensive tutorial paper on coronal MHD seismology \cite{2024RvMPP...8...19N}, while Queen's University Belfast recently commissioned the first spectropolarimetric solar IFU designed for the blue portion of the optical spectrum \cite{2023SoPh..298..146J} and hosts a dedicated wave analysis tools coding repository \cite{2025NRvMP...5...21J} (\href{https://WaLSA.tools}{https://WaLSA.tools}) to assist researchers around the globe accurately undertake seismological investigations. Northumbria and Sheffield and Warwick have significantly advanced MHD wave theories. Substantial research groups in both observational and theoretical work in the terrestrial context  exist at, e.g., Leicester, Imperial College, Northumbria, and St Andrews, with work in a broader planetary context centred at Leicester, Imperial College, and UCL. MHD forms an important strand within the research programmes of major current UK investments, such as SuperDARN, EISCAT~3D, BepiColombo, and JUICE.

~ \\

\vspace{-5mm}
\noindent\rule[7pt]{\linewidth}{0.4pt}

\vspace{-8.3mm}
\begin{snugshade*}
\begin{center}
{\bfseries Partnership opportunities} 
\end{center}
\end{snugshade*}

\vspace{-5.3mm}
\noindent\rule[7pt]{\linewidth}{0.4pt}

\vspace{-3mm}
Efforts and achievements in MHD seismology create a solid ground for a number of partnership opportunities in solar, heliospheric, geophysical, and stellar research. It includes involvement and leadership in bilateral and multi-lateral projects. For example, UK actively participates in major ongoing international solar projects, such as SKA, Solar-C, and DKIST, plus has existing partnerships with SuperDARN, SMILE, EISCAT~3D, BepiColombo, JUICE, and JWST.

\hspace{5mm} Furthermore, members of the UK community sit on the Science Boards and Science Advisory Groups of upcoming ground-based and space-borne facilities, such as the European Solar Telescope (EST), Indian National Large Solar Telescope (NLST), and the Brazilian Galileo Solar Space Telescope (GSST). This involvement is critical and enables members of the UK scientific community to shape and refine the objectives and observing strategies of these future missions. Furthermore, close collaborations with such high-profile international institutes will provide extra opportunities for the UK to participate in instrument design/assembly, bilateral funding calls, and provide global employment opportunities for early career researchers across the UK. 

\newpage
\setcounter{page}{1}
\thispagestyle{empty}
{\Large{\textbf{List of Signatories}}}

\begin{table}[!ht]
    \centering
    \begin{tabular}{|l|l|l|}
    \hline
        \textbf{Signatory Name} & \textbf{Affiliation} & \textbf{Country} \\ \hline
        Viktor Fedun & Sheffield University & UK \\ \hline
        Istvan Ballai & Sheffield University & UK \\ \hline
        Jinge Zhang & Paris Observatory - PSL & France \\ \hline
        Samuel Skirvin & Northumbria University & UK \\ \hline
        Mihalis Mathioudakis & Queen's University Belfast & UK \\ \hline
        Anshu Kumari & Physical Research Laboratory, India & India \\ \hline
        Divya Paliwal & Physical Research Laboratory, India & India \\ \hline
        David Jess & Queen's University Belfast & UK \\ \hline
        Shahin Jafarzadeh & Queen's University Belfast & UK \\ \hline
        Glen Chambers & Queen's University Belfast & UK \\ \hline
        Samuel Grant & Queen's University Belfast & UK \\ \hline
        Luis Eduardo Antunes Vieira & Instituto Nacional de Pesquisas Espaciais & Brazil \\ \hline
        Michele Berretti & Università di Trento / Università degli Studi di Roma & Italy \\ \hline
        Abhirup Datta & Indian Institute of Technology Indore & India \\ \hline
        Shreeyesh Biswal & University of Sheffield & UK \\ \hline
        Michael Ruderman & University of Sheffield & UK \\ \hline
        Szabolcs Soos & Eotvos University & Hungary \\ \hline
        Sahel Dey & Newcastle University & Australia \\ \hline
        Alexander Nindos & University of Ioannina & Greece \\ \hline
        Spiros Patsourakos & University of Ioannina & Greece \\ \hline
        Kris Mrwaski & University of Marie Curie Sklodowska & Poland \\ \hline
        Piyali Chatterjee & Indian Institute of Astrophysics & India \\ \hline
        Marianna Korsos & University of Sheffield & UK \\ \hline
        Maria Madjarska & Max-Planck Institute for Solar System Research & Germany~~ \\ \hline
        Paul Cally & Monash University & Australia \\ \hline
        Dibyendu Nandi & IIESR Kolkotta & India \\ \hline
        Dipankar Banerjee & Indian Institute of Space Sciecne and Technology & India \\ \hline
        Dyrgesh Tripathi & IUCAA Pune & India \\ \hline
        Jiajia Liu & USTC Hefei & China \\ \hline
        Daria Shukhoboskaia & Royal Observatory of Belgium & Belgium \\ \hline
        Tom Van Doorsselaere & K.U. Leuven & Belgium \\ \hline
        Ricardo Gafiera & University of Coimbra & Portugal \\ \hline
        Marco Stangalini & Italian Space Agency & Italy \\ \hline
        Dario Del Moro & University of Rome Tor Vergata & Italy \\ \hline
        Stefaan Poedts & K.U. Leuven & Belgium \\ \hline
        Lei Ni & Yunnan Observatories of CAS & China \\ \hline
        Tim Yeoman & University of Leicester & UK \\ \hline
        James McLaughlin & Northumbria University & UK \\ \hline
        Sergey Belov & University of Warwick & UK \\ \hline
        Dan Ratliff & Northumbria University & UK \\ \hline
        Sarah Bentley & Northumbria University & UK \\ \hline
        Tom Elsden & University of St Andrews & UK \\ \hline
        Andrew Wright & University of St Andrews & UK \\ \hline
        Ryan Smith & Northumbria University & UK \\ \hline
        Gert Botha & Northumbria University & UK \\ \hline
        Damian Christian & California State University Northridge (CSUN) & USA \\ \hline
        Thomas Rees-Crockford & Northumbria University & UK \\ \hline
        Cristina Cadavid & California State University Northridge (CSUN) & USA \\ \hline
     \end{tabular}
\end{table}

\newpage
\setcounter{page}{1}
\thispagestyle{empty}
\begin{table}[!ht]
    \centering
    \begin{tabular}{|l|l|l|}
    \hline
        \textbf{Signatory Name} & \textbf{Affiliation} & \textbf{Country} \\ \hline
        Manuel Collados & IAC & Spain \\ \hline
        Elena Khomenko & IAC & Spain \\ \hline
        Balazs Asztalos & Eotvos University & Hungary~~~~ \\ \hline
        Rebecca Meadowcroft & University of Warwick & UK \\ \hline
        Ben Snow & University of Exeter & UK \\ \hline
        Fernando Luis Guarnieri~~~~~~~ & Instituto Nacional de Pesquisas Espaciais~~~~~~~~~~~~~~~ & Brazil \\ \hline
        Alisson Dal Lago & Instituto Nacional de Pesquisas Espaciais & Brazil \\ \hline
        Paulo Jauer & Instituto Nacional de Pesquisas Espaciais & Brazil \\ \hline
        Franciele Carlesso & Instituto Nacional de Pesquisas Espaciais & Brazil \\ \hline
        Alex Degeling & Shandong University & China \\ \hline
        Gary Verth & University of Sheffield & UK \\ \hline
        Jasmine Sandhu & University of Leicester & UK \\ \hline    
        Yash Saneshwar & Northumbria University & UK \\ \hline
        Eamon Scullion & Northumbria University & UK \\ \hline
        Hamish Reid & MSSL - UCL & UK \\ \hline
        Keshav Aggarwal & Indian Institute of Technology Indore & India \\ \hline
        Ian Mann & University of Alberta & Canada \\ \hline
    \end{tabular}
\end{table}

\renewcommand{\thepage}{}
\bibliographystyle{unsrtnat}
\bibliography{references}

\end{document}